\documentclass[preprint,nofootinbib]{revtex4}
\usepackage{amsfonts}
\usepackage{amsmath}
\usepackage{graphicx}

\begin{document}

\title{Static wormhole solution for higher-dimensional gravity in vacuum}
\author{Gustavo Dotti$^{1}$, Julio Oliva$^{2,3}$, and Ricardo Troncoso$^{3}$}
\affiliation{$^{1}$Facultad de Matem\'{a}tica, Astronom\'{\i}a y
F\'{\i}sica, Universidad Nacional de C\'{o}rdoba, Ciudad
Universitaria, (5000) C\'{o}rdoba, Argentina. \\
$^{2}$Departamento de F\'{\i}sica, Universidad de Concepci\'{o}n,
Casilla, 160-C, Concepci\'{o}n, Chile. \\ $^{3}$Centro de Estudios
Cient\'{\i}ficos (CECS), Casilla 1469, Valdivia, Chile.}

\begin{abstract}
A static wormhole solution for gravity in vacuum is found for odd dimensions
greater than four. In five dimensions the gravitational theory considered is
described by the Einstein-Gauss-Bonnet action where the coupling of the
quadratic term is fixed in terms of the cosmological constant. In higher
dimensions $d=2n+1$, the theory corresponds to a particular case of the
Lovelock action containing higher powers of the curvature, so that in
general, it can be written as a Chern-Simons form for the AdS\ group. The
wormhole connects two asymptotically locally AdS spacetimes each with a
geometry at the boundary locally given by $\mathbb{R}\times S^{1}\times
H_{d-3}$. Gravity pulls towards a fixed hypersurface located at some
arbitrary proper distance parallel to the neck. The causal structure shows
that both asymptotic regions are connected by light signals in a finite
time. The Euclidean continuation of the wormhole is smooth independently of
the Euclidean time period, and it can be seen as instanton with vanishing
Euclidean action. The mass can also be obtained from a surface integral and
it is shown to vanish.
\end{abstract}

\maketitle

\section{Introduction}
The quest for exact wormhole solutions in General Relativity, which are
handles in the spacetime topology, has appeared repeatedly in theoretical
physics within different subjects, ranging from the attempt of describing
physics as pure geometry, as in the ancient Einstein-Rosen bridge model of a
particle \cite{Einstein-Rosen}, to the concept of \emph{\textquotedblleft
charge without charge" }\cite{Wheeler-charge}, as well as in several issues
concerning the Euclidean approach to quantum gravity (see, e.g., \cite%
{Gibbons Hawking}). More recently, during the 80's, motivated by the
possibility of quick interstellar travel, Morris, Thorne and Yurtsever
pushed forward the study of wormholes from the point of view of
\textquotedblleft reverse engineering", i.e., devising a suitable geometry
that allows this possibility, and making use of the Einstein field equations
in order to find the corresponding stress-energy tensor that supports it as
an exact solution \cite{MTY}. However, one of the obstacles to circumvent,
for practical affairs, is the need of exotic forms of matter, since it is
known that the required stress-energy tensor does not satisfy the standard
energy conditions (see, e.g., \cite{Visser}). Besides, the pursuit of a
consistent framework for a unifying theory of matter and interactions has
led to a consensus in the high energy community that it should be formulated
in dimensions higher than four. However, for General Relativity in higher
dimensions, the obstacle aforementioned concerning the stress-energy tensor
persists. Nonetheless, in higher dimensions, the straightforward dimensional
continuation of Einstein's theory is not the only option to describe
gravity. Indeed, even from a conservative point of view, and following the
same basic principles of General Relativity, the most general theory of
gravity in higher dimensions that leads to second order field equations for
the metric is described by the Lovelock action \cite{Lovelock}, which is
nonlinear in the curvature. In this vein, for the simplest extension, being
quadratic in the curvature, it has been found that the so-called
Einstein-Gauss-Bonnet theory, admits wormhole solutions that would not
violate the weak energy condition provided the Gauss-Bonnet coupling
constant is negative and bounded according to the shape of the solution \cite%
{Bhawal-Kar}.

Here it is shown that in five dimensions, allowing a cosmological (volume)
term in the Einstein-Gauss-Bonnet action, and choosing the coupling constant
of the quadratic term such that the theory admits a single anti-de Sitter
(AdS) vacuum, allows the existence of an exact static wormhole solution in
vacuum. As explained below, the solution turns out to have \textquotedblleft
mass without mass" and connects two asymptotically locally AdS spacetimes
each with a geometry at the boundary that is not spherically symmetric. It
is worth to remark that no energy conditions can be violated since the whole
spacetime is devoid of any kind of stress-energy tensor. In what follows,
the five-dimensional case is worked out in detail, and next we explain how
the results extend to higher odd dimensions for a special class of theories
among the Lovelock class, which are also selected by demanding the existence
of a unique AdS vacuum.

\section{Static wormhole in five dimensions}
The action for the
Einstein-Gauss-Bonnet theory with a volume term can be written as%
\begin{equation*}
I_{5}=\kappa \!\int \!\epsilon _{abcde}\left( \!R^{ab}R^{cd}\!+\!\frac{2}{%
3l^{2}}R^{ab}e^{c}e^{d}\!+\!\frac{1}{5l^{4}}e^{a}e^{b}e^{c}e^{d}\!\right)
\!e^{e},
\end{equation*}%
where $R^{ab}=d\omega ^{ab}+\omega _{\text{ \ }f}^{a}\omega ^{fb}$ is the
curvature $2$-form for the spin connection $\omega ^{ab}$, and $e^{a}$ is
the vielbein. The coupling of the Gauss-Bonnet term has been fixed so that
the theory possesses a unique AdS vacuum of radius $l$. In the absence of
torsion, the field equations can be simply written as%
\begin{equation}
\mathcal{E}_{a}:\mathcal{=}\epsilon _{abcde}\bar{R}^{bc}\bar{R}^{de}=0,
\label{feqfive}
\end{equation}%
where $\bar{R}^{bc}:=R^{bc}+\frac{1}{l^{2}}e^{b}e^{c}$. These equations are
solved by the following metric%
\begin{equation}
ds_{5}^{2}\!=\!l^{2}\!\left[ -\cosh ^{2}\left( \rho -\rho _{0}\right)
dt^{2}\!+d\rho ^{2}\!+\cosh ^{2}\left( \rho \right) d\Sigma _{3}^{2}\right]
\!,  \label{metric}
\end{equation}%
where $\rho _{0}$ is an integration constant and $d\Sigma _{3}^{2}$ stands
for the metric of the base manifold which can be chosen to be locally of the
form $\Sigma _{3}=S^{1}\times H_{2}$. The radius of the hyperbolic manifold $%
H_{2}$ turns out to be $3^{-1/2}$, so that the Ricci scalar of $\Sigma _{3}$
has the value of $-6$, as required by the field equations. The metric (\ref%
{metric}) describes a static wormhole with a neck of radius $l$, located at
the minimum of the warp factor of the base manifold, at $\rho =0$. Since $%
-\infty <\rho <\infty $, the wormhole connects two asymptotically locally
AdS spacetimes so that the geometry at the boundary is locally given by $%
\mathbb{R}\times S^{1}\times H_{2}$. Actually, it is simple to check that
the field equations are solved provided the base manifold $\Sigma _{3}$ has
a negative constant Ricci scalar. Indeed, for a metric of the form (\ref%
{metric}) the vielbeins can be chosen as%
\begin{equation*}
e^{0}=l\ \cosh (\rho -\rho _{0})dt\ ;\ e^{1}=l\ d\rho \ ;\ e^{m}=l\cosh
(\rho )\ \tilde{e}^{m}\ ,
\end{equation*}%
where $\tilde{e}^{m}$ is the dreibein of $\Sigma _{3}$, so that curvature
two-form is such that the only nonvanishing components of $\bar{R}^{ab}$ read%
\begin{equation}
\bar{R}^{0m}=\cosh \left( \rho _{0}\right) dt\wedge \tilde{e}^{m}\ ;\ \bar{R}%
^{mn}=\tilde{R}^{mn}+\tilde{e}^{m}\tilde{e}^{n}~.  \label{CurvatureBar}
\end{equation}%
Replacing Eqs. (\ref{CurvatureBar}) in the field equations (\ref{feqfive})
it turns out that the components $\mathcal{E}_{0}$ and $\mathcal{E}_{m}$ are
identically satisfied. The remaining field equation, $\mathcal{E}_{1}=0$,
reads%
\begin{equation*}
\cosh \left( \rho _{0}\right) dt\wedge \epsilon _{mnp}\left( \tilde{R}^{mn}%
\tilde{e}^{p}+\tilde{e}^{m}\tilde{e}^{n}\tilde{e}^{p}\right) =0\ ,
\end{equation*}%
which implies that $\Sigma _{3}$ must be a manifold with a constant Ricci
scalar satisfying%
\begin{equation}
\tilde{R}=-6\ .  \label{RicciScalar5}
\end{equation}%
One can notice that the field equations (\ref{feqfive}) are deterministic
unless $\Sigma _{3}$ were chosen as having negative constant curvature,
i.e., being locally isomorphic to $H_{3}$. If so, the field equations would
degenerate in such a way that the component $g_{tt}$ of the metric becomes
an arbitrary function of $\rho $. This degeneracy is a known feature of the
class of theories considered here \cite{Ms-T-Z}, and is overcome by choosing
a base manifold satisfying (\ref{RicciScalar5}) but not being of constant
curvature. A simple example of a compact smooth three-dimensional manifold
fulfilling these conditions is given by $\Sigma _{3}=S^{1}\times
H_{2}/\Gamma $, where $H_{2}$ has radius $\frac{1}{\sqrt{3}}$, and $\Gamma $
is a freely acting discrete subgroup of $O(2,1)$. It worth pointing out that
$\Sigma _{3}$ is not an Einstein manifold, and that any nontrivial solution
of the corresponding Yamabe problem (see e.g. \cite{Yamabe}) provides a
suitable choice for $\Sigma _{3}$.

The causal structure of the wormhole is depicted in Fig. 1, where the dotted
vertical line shows the position of the neck, and the solid bold lines
correspond to the asymptotic regions located at $\rho =\pm \infty $, each of
them resembling an AdS spacetime but with a different base manifold since
the usual sphere $S^{3}$ must be replaced by $\Sigma _{3}$. The line at the
center stands for $\rho =\rho _{0}$.
\begin{figure}[tbp]
\includegraphics{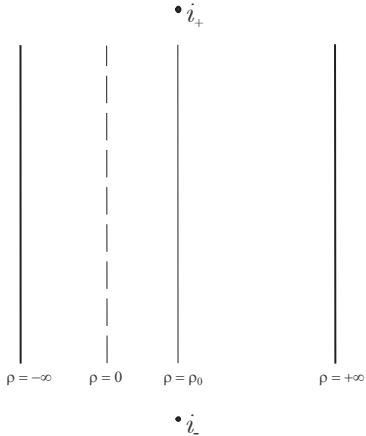}
\caption{Penrose diagram for the wormhole}
\label{fig:epsart}
\end{figure}
It is apparent from the diagram that null and timelike curves can go forth
and back from the neck. Furthermore, note that radial null geodesics are
able to connect both asymptotic regions in finite time. Indeed, one can see
from (\ref{metric}) that the coordinate time that a photon takes to travel
radially from one asymptotic region, $\rho =-\infty $, to the other at $\rho
=+\infty $ is given by%
\begin{equation*}
\Delta t=\int_{-\infty }^{+\infty }\frac{d\rho }{\cosh \left( \rho -\rho
_{0}\right) }=\left[ 2\arctan \left( e^{\rho -\rho _{0}}\right) \right]
_{-\infty }^{+\infty }=\pi ,
\end{equation*}%
which does not depend on $\rho _{0}$. Thus, any static observer located at $%
\rho =\rho _{0}$ says that this occurred in a proper time given by $\pi l$.
Note also that this observer actually lives on a static timelike geodesic,
and it is easy to see that a small perturbation along $\rho $ makes him to
oscillate around $\rho =\rho _{0}$. This means that gravity is pulling
towards the fixed hypersurface defined by $\rho =\rho _{0}$ which is
parallel to the neck. Hence, the constant $\rho _{0}$ corresponds to a
modulus parametrizing the proper distance between this hypersurface and the
neck. Actually, one can explicitly check that radial timelike geodesics are
always confined since they satisfy%
\begin{align*}
\frac{1}{2}\dot{\rho}^{2}-\frac{E^{2}}{2\cosh ^{2}\left( \rho -\rho
_{0}\right) }& =C_{0}, \\
\dot{t}-\frac{E}{\cosh ^{2}\left( \rho -\rho _{0}\right) }& =0,
\end{align*}%
where the dot stand for derivatives with respect to the proper time $\tau $,
and the velocity is normalized as $u_{\mu }u^{\mu }=2l^{2}C_{0}$. Thus, one
concludes that the position of a radial geodesic, $\rho (\tau )$, in proper
time behaves as a particle in a P\"{o}schl-Teller potential. Therefore, as
it can also be seen from the Penrose diagram, null and spacelike radial
geodesics connect both asymptotic regions in finite time. Furthermore,
timelike geodesics for which $2l^{2}C_{0}=-1$ are shown to be confined.

\subsection{Euclidean continuation and a finite action principle}
The
Euclidean continuation of the wormhole metric (\ref{metric}) is smooth
independently of the Euclidean time period, so that the wormhole could be in
thermal equilibrium with a heat bath of arbitrary temperature. It is then
useful to evaluate the Euclidean action for this configuration. It has been
shown in \cite{MOTZ} that the action $I_{5}$ can be regularized by adding a
suitable boundary term in a background independent way, which can be written
just in terms of the extrinsic curvature and the geometry at the boundary.
The total action then reads $I_{T}=I_{5}-B_{4}$, where the boundary term
reads%
\begin{equation}
B_{4}\!=\!\kappa \!\int_{\partial M}\!\epsilon _{abcde}\theta
^{ab}e^{c}\left( \!R^{de}-\frac{1}{2}\theta _{\ f}^{d}\theta ^{fe}+\frac{1}{%
6l^{2}}e^{d}e^{e}\!\right) .  \label{BoundaryTerm5}
\end{equation}%
For the wormhole solution, the boundary of $M$ is of the form, $\partial
M=\partial M_{+}\cup \partial M_{-}$, where $\partial M_{-}$ has a reversed
orientation with respect to that of $\partial M_{+}$. Using the fact that
the only non vanishing components of the second fundamental form $\theta
^{ab}$ for the wormhole (\ref{metric}), for each boundary, are given by%
\begin{equation}
\theta ^{01}=\sinh \left( \rho -\rho _{0}\right) d\tau ~;~\theta ^{m1}=\sinh
\left( \rho \right) \tilde{e}^{m}~,  \label{thetas}
\end{equation}%
where $\tau $ now stands for the Euclidean time. It is simple to verify that
the action principle $I_{T}$ attains an extremum for the wormhole solution.

Let us evaluate the action $I_{T}$ for the wormhole (\ref{metric}) with a
base manifold of the form $\Sigma _{3}=S^{1}\times H_{2}/\Gamma $. Assuming
that the boundaries, are located at $\rho =\rho _{+}$ and $\rho =\rho _{-}$,
respectively, one obtains that%
\begin{equation*}
I_{5}\!=\!B_{4}\!=\!2\kappa l\beta \sigma \left[ 3\sinh \left( \rho
_{0}\right) \!+\!8\cosh ^{3}\left( \rho \right) \sinh \left( \rho \!-\!\rho
_{0}\right) \right] _{\rho _{-}}^{\rho _{+}},
\end{equation*}%
where $\beta $ is the Euclidean time period, and $\sigma =\frac{8\pi ^{2}}{3}%
R_{0}(g-1)$ is the volume of the base manifold $\Sigma _{3}$ in terms of the
radius of $S^{1}$ and the genus of $H_{2}/\Gamma $, given by $R_{0}$ and $g$%
, respectively.

Therefore, remarkably, the regularized Euclidean action $I_{T}$ does not
depend on the integration constant $\rho _{0}$, and it vanishes for each
boundary regardless their position. This means that the Euclidean
continuation of the wormhole can be seen as an instanton with vanishing
Euclidean action. Consequently, the total mass of the wormhole is found to
vanish since $M=-\frac{\partial I_{T}}{\partial \beta }=0$. The same results
extend to any base manifold with a Ricci scalar satisfying Eq. (\ref%
{RicciScalar5}).

It is worth pointing out that the value of the regularized action for the
wormhole is lower than the one for AdS spacetime, which turns out to be $%
I_{T}\left( \mathrm{AdS}\right) =6\Omega _{3}\kappa \beta $, where $\Omega
_{3}$ is the volume of $S^{3}$. However, AdS spacetime has a negative
\textquotedblleft vacuum energy" given by $M_{AdS}=-6\Omega _{3}\kappa $.

\subsection{Mass from a surface integral}
The fact that the action principle $%
I_{T}$ has an extremum for the wormhole solution, also allows to compute the
mass from the following surface integral \cite{MOTZ}%
\begin{equation}
Q\left( \xi \right) =\kappa \!\int_{\partial \Sigma }\!\epsilon \left(
I_{\xi }\theta \frac{e}{l}+\theta I_{\xi }\frac{e}{l}\right) \!\left( \!%
\tilde{R}+\frac{1}{2}\theta ^{2}+\frac{1}{2l^{2}}e^{2}\right) ,  \label{Qchi}
\end{equation}%
which is obtained by the straightforward application of Noether's theorem
\footnote{In order to simplify the notation, tangent space
indices are assumed to be contracted in canonical order. The action of the
contraction operator $I_{\xi }$ over a $p$-form $\alpha _{p}=\frac{1}{p!}%
\alpha _{\mu _{1}\cdots \mu _{p}}dx^{\mu _{1}}\cdots dx^{\mu _{p}}$ is given
by $I_{\xi }\alpha _{p}=\frac{1}{(p-1)!}\xi ^{\nu }\alpha _{\nu \mu
_{1}\cdots \mu _{p-1}}dx^{\mu _{1}}\cdots dx^{\mu _{p-1}}$, and $\partial
\Sigma $ stands for the boundary of the spacelike section.}. The mass is obtained evaluating (\ref{Qchi}) for the
timelike Killing vector $\xi =\partial _{t}$, and one then confirms that the
mass, $M=Q\left( \partial _{t}\right) $, vanishes for the Lorenzian
solution. We would like to remark that, following this procedure, one
obtains that the contribution to the total mass coming from each boundary
reads%
\begin{equation}
Q_{\pm }\left( \partial _{t}\right) =\pm 6\sigma \kappa \sinh \left( \rho
_{0}\right) ,
\end{equation}%
where $Q_{\pm }\left( \partial _{t}\right) $ is the value of (\ref{Qchi}) at
$\partial \Sigma _{\pm }$, which again does not depend on $\rho _{+}$ and $%
\rho _{-}$. This means that for a positive value of $\rho _{0}$, the mass of
the wormhole appears to be positive for observers located at $\rho _{+}$,
and negative for the ones at $\rho _{-}$, such that the total mass always
vanishes. This provides a concrete example of what Wheeler dubbed \emph{%
\textquotedblleft mass without mass"}. Hence, the integration constant $\rho
_{0}$ could also be regarded as a parameter for the apparent mass at each
side of the wormhole, which vanishes only when the solution acquires
reflection symmetry, i.e., for $\rho _{0}=0$.

\section{The wormhole in higher odd dimensions}
The five-dimensional
static wormhole solution in vacuum, given by Eq. (\ref{metric}), can be
extended as an exact solution for a very special class of gravity theories
among the Lovelock family in higher odd dimensions $d=2n+1$. In analogy with
the procedure in five dimensions, the theory can be constructed so that the
relative couplings between each Lovelock term are chosen so that the action
has the highest possible power in the curvature and possesses a unique AdS
vacuum of radius $l$. The field equations then read%
\begin{equation}
\mathcal{E}_{A}:\mathcal{=}\epsilon _{ab_{1}\cdot \cdot \cdot b_{2n}}\bar{R}%
^{b_{1}b_{2}}\cdot \cdot \cdot \bar{R}^{b_{2n-1}b_{2n}}=0,  \label{feqD}
\end{equation}%
which are solved by the straightforward extension of (\ref{metric}) to
higher dimensions%
\begin{equation*}
ds^{2}\!=\!l^{2}\!\left[ -\cosh ^{2}\left( \rho -\rho _{0}\right)
dt^{2}\!+d\rho ^{2}\!+\cosh ^{2}\left( \rho \right) d\Sigma _{2n-1}^{2}%
\right] ,
\end{equation*}%
where $\rho _{0}$ is an integration constant, and $d\Sigma _{2n-1}^{2}$
stands for the metric of the base manifold. In the generic case, the base
manifold must solve the following equation \footnote{Note that this equation corresponds to the trace of the
Euclidean field equations for the same theory in $2n-1$ dimensions with a
unit AdS radius.}%
\begin{equation}
\epsilon _{m_{1}\cdot \cdot \cdot m_{2n-1}}\bar{R}^{m_{1}m_{2}}\cdot \cdot
\cdot \bar{R}^{m_{2n-3}m_{2n-2}}\tilde{e}^{m_{2n-1}}=0,  \label{TrCS}
\end{equation}%
where $\tilde{e}^{m}$ is the vielbein of $\Sigma _{_{2n-1}}$. Note that this
is a single scalar equation.

As in the five-dimensional case, it is worth to remark that the field
equations (\ref{feqD}) are deterministic unless $\Sigma _{_{2n-1}}$ solves
the field equations for the same theory in $2n-1$ dimensions with unit AdS
radius. If so, the field equations would degenerate and the metric component
$g_{tt}$ would be an arbitrary function of $\rho $. In particular, the
hyperbolic space $H_{2n-1}$ falls within this degenerate class. Therefore,
in order to circumvent this degeneracy, the base manifold must fulfill Eq. (%
\ref{TrCS}), but without solving simultaneously the field equations for the
same theory in $2n-1$ dimensions with unit AdS radius.

A\ simple example of a compact smooth $(2n-1)$-dimensional manifold
fulfilling the latter conditions is given by $\Sigma _{2n-1}=S^{1}\times
H_{2n-2}/\Gamma $, where $H_{2n-2}$ has radius $(2n-1)^{-1/2}$, and $\Gamma $
is a freely acting discrete subgroup of $O(2n-2,1)$. Note that $\Sigma
_{2n-1}$ is not an Einstein manifold.

The metric in higher odd dimensions then describes a static wormhole with a
neck of radius $l$ connecting two asymptotic regions which are locally AdS
spacetimes, so that the geometry at the boundary is given by $\mathbb{R}%
\times S^{1}\times H_{2n-2}/\Gamma $. The wormhole in higher dimensions
shares the features described in the five-dimensional case, including the
meaning of the parameter $\rho _{0}$, and its causal structure is depicted
in Fig. 1.

As in the five-dimensional case, the Euclidean continuation of the wormhole
metric is smooth and it has an arbitrary Euclidean time period. The
Euclidean action can be regularized in higher odd dimensions in a background
independent way as in Ref. \cite{MOTZ}, by the addition of a suitable
boundary term which is the analogue of (\ref{BoundaryTerm5}), and can also
be written in terms of the extrinsic curvature and the geometry at the
boundary. The nonvanishing components of the second fundamental form $\theta
^{ab}$ acquire the same form as in Eq. (\ref{thetas}) for higher dimensions,
so that it is easy to check that the regularized action has an extremum for
the wormhole solution. As in the five-dimensional case, the Euclidean
continuation of the wormhole can be seen as an instanton with a regularized
action that vanishes independently of the position of the boundaries, so
that its mass is also found to vanish. This means that AdS spacetime has a
greater action than the wormhole, but a lower \textquotedblleft vacuum
energy".

The wormhole mass for the Lorenzian solution can also be shown to vanish
making use of a surface integral which is the extension of (\ref{Qchi}) to
higher odd dimensions \cite{MOTZ}. The contribution to the total mass coming
from each boundary does not depend on the location of the boundaries and is
given by%
\begin{equation*}
Q_{\pm }\left( \partial _{t}\right) =\pm \alpha _{n}\sigma \kappa \sinh
\left( \rho _{0}\right) ,
\end{equation*}%
so that for a nonvanishing integration constant $\rho _{0}$, the wormhole
appears to have \textquotedblleft mass without mass". Here $\alpha _{n}:=%
\left[ (1-2n)^{n-1}-2^{n}(1-n)^{n-1}\right] (2n-1)!$.

It is simple to show that for different base manifolds, the Euclidean action
also vanishes, and the surface integrals for the mass possess a similar
behavior.

\section{Final remarks}
 The existence of interesting solutions in vacuum
could be regarded as a criterion to discriminate among the different
possible gravity theories that arise only in dimensions greater than four.
Indeed, it has been shown that, among the Lovelock family, selecting the
theories as having a unique maximally symmetric vacuum solution guarantees
the existence of well-behaved black hole solutions \cite{BH-Scan}, \cite{ATZ}%
. In turn, it has been shown that demanding the existence of simple
compactifications describing exact black $p$-brane solutions, selects the
same class of theories \cite{GOT} (see also \cite{Kastor-Mann}). In this
sense, the theory possessing the highest possible power in the curvature
with a unique AdS vacuum is particularly interesting and it is singled out
for diverse reasons. It is worth to mention that in this case, the
Lagrangian can be written as a Chern-Simons gauge theory for the AdS\ group
\cite{Chamseddine}, so that the local symmetry is enlarged from Lorentz to
AdS. One can see that the wormhole solution found here is not only a vacuum
solution for these theories, but also for their locally supersymmetric
extension in five \cite{Chamseddine2} and higher odd dimensions \cite{TZ}.
The compactification of the wormhole solution is straightforward since it
has been shown that it always admits a base manifold with a $S^{1}$ factor.
This means that in one dimension below, the geometrical and causal behavior
is similar to the one described here, but in this case the base manifold is
allowed to be locally a hyperbolic space without producing a degeneracy of
the field equations. Note that the dimensionally reduced solution is
supported by a nontrivial dilaton field with a nonvanishing stress-energy
tensor.

\textit{Acknowledgments.-- }We thank S. Willison for helpful discussions.
G.D is supported by CONICET. J.O. thanks the project MECESUP UCO-0209. This
work was partially funded by FONDECYT grants 1040921, 1051056, 1061291;
Secyt-UNC and CONICET. The generous support to Centro de Estudios Cient\'{\i}%
ficos (CECS) by Empresas CMPC is also acknowledged. CECS is funded in part
by grants from the Millennium Science Initiative, Fundaci\'{o}n Andes and
the Tinker Foundation.

\end{document}